\documentclass[reprint,superscriptaddress]{revtex4-1}
\usepackage{subscript}
\usepackage{graphicx}
\usepackage{amsmath}
\usepackage{multirow}
\usepackage{lineno}

\begin{document}

\preprint{}

\title{Antineutrino Monitoring of Thorium Reactors}


\author{Oluwatomi A. Akindele}
\email[]{Correspondence to: University of California,4155 Etcheverry Hall, MC 1730, Berkeley,CA 94720-1730; Email: tomi90@berkeley.edu}
\affiliation{Department of Nuclear Engineering, University of California, Berkeley, California 94720, USA}
\author{Adam Bernstein}
\affiliation{Lawrence Livermore National Laboratory, Livermore, California 94550, USA}

\author{Eric B. Norman}
\affiliation{Department of Nuclear Engineering, University of California, Berkeley, California 94720, USA}

\date{\today}

\begin{abstract}
Various groups have demonstrated that antineutrino monitoring can
be successful in assessing the plutonium content in water-cooled nuclear
reactors for nonproliferation applications. New reactor designs and
concepts incorporate nontraditional fuels types and chemistry. Understanding
how these properties affect the antineutrino emission from a reactor
can extend the applicability of antineutrino monitoring. Thorium molten
salt reactors (MSR) breed \textsuperscript{233}U, that if diverted
constitute a direct use material as defined by the International Atomic Energy Agency (IAEA). The antineutrino spectrum
from the fission of \textsuperscript{233}U has been estimated for the first time, and the feasibility
of detecting the diversion of 8 kg of \textsuperscript{233}U,
within a 30 day timeliness goal has been evaluated. The antineutrino
emission from a thorium reactor operating under normal conditions
is compared to a diversion scenario by evaluating
the daily antineutrino count rate and the energy spectrum of the detected
antineutrinos at a 25 meter standoff. It was found that the diversion of a significant quantity
of \textsuperscript{233}U could not be detected within the current IAEA timeliness
detection goal using either tests. A rate-time based analysis exceeded the timeliness
goal by 23 days, while a spectral based analysis exceeds this goal by 31 days.

\end{abstract}


\maketitle

\section{Introduction}
The accident that occurred at the Fukushima-Daiichi nuclear power plant
following the 2011 earthquake and subsequent tsunami in Japan led
to an international effort to increase the development of accident
tolerant nuclear fuels specifically to withstand a core meltdown in
the case of a beyond-design-basis event such as a loss of coolant.
This has led to a revival in the interest  in thorium molten salt reactor (MSR) designs.
Despite the improvement to reactor safety, the production and online
reprocessing of \textsuperscript{233}Pa which decays into
\textsuperscript{233}U creates a state-sponsored proliferation risk.
Current IAEA methods used to
detect the divergence of nuclear material are relatively intrusive
and cannot directly measure the fissile inventory of the reactor \cite{1}. 
Moreover, the safeguards approach for this new class of reactors has not yet been defined.
Groups at Lawrence Livermore National Laboratory (LLNL) and Sandia National
Laboratories (SNL) have shown that using antineutrino detectors provide
a remote and non-intrusive method to observe the diversion of nuclear
material \cite{2}. In this paper the feasibility of using antineutrino
monitoring to detect the diversion of \textsuperscript{233}U from a thorium MSR is evaluated. 

\section{Background}

Antineutrinos are weakly interacting neutral particles produced from
the beta decay of neutron rich fission products. Unlike neutrons,
beta particles, and gamma rays; antineutrinos can be detected from well
outside the reactor core because of their low interaction cross section.
About six antineutrinos are produced from each fission resulting in
an antineutrino flux of about 10\textsuperscript{21}$ \overline{\nu_{e}}$/s
from a 1 GW reactor. Antineutrino monitoring of traditional nuclear
reactors uses the varying spectral contributions of fission products
from \textsuperscript{235}U and \textsuperscript{239}Pu over time
to make the detection of materials diversion feasible. When these
isotopes fission they produce different beta emitters with various yields.
The antineutrino spectrum for each fissile
isotope is an combination of the antineutrino spectra of individual fission products, with each contribution fixed by
the fission yield and branching ratio. As a result, each fissile
isotope has a distinct antineutrino spectra shape and amplitude making
its growth or depletion potentially accessible through antineutrino monitoring. For
thorium reactors the difference in antineutrino spectra from \textsuperscript{233}U
and \textsuperscript{235}U, the most prominent fissile isotopes,
allows for diversion detection.

\subsection{Antineutrino Detection}

Antineutrinos can be detected through their interactions with quasi-free protons in a liquid
scintillating medium, using the inverse beta decay process shown in Eq. 1.
\begin{equation}
\overline{\nu}_{e}+p\,\longrightarrow e^{+}+n\quad Q=-1.804\, MeV
\end{equation}
The threshold for this reaction is 1.804 MeV, limiting detection
of antineutrinos to those above that energy. In a liquid scintillator detector, the positron created from
this reaction will slow in the detector and annihilate with an electron.
The positron will deposit almost all of its energy at the end of its
trajectory, and because subsequent annihilation occurs instantaneously,
the energy of the positron and the gamma rays emitted from annihilation
create an indistinguishable prompt signal. In a gadolinium-doped scintillator, the neutron captures
on the Gd dopant within a few tens of microseconds, depending on the dopant concentration, a delayed signal from the resulting
8 MeV gamma ray cascade is formed. These two signals, detected in close time
coincidence defines an antineutrino event. 

In the present work, the antineutrino emission from the reactor is studied using a design similiar to that of a detector
developed and tested at LLNL \cite{classen}. The detector is assumed deployed at a 25 meter standoff, 10 meters below the surface, equivalent to a water 
overburden of about 25 meters. At these depths a previous iteration of this detector, SONGS1, experienced a singles and delayed coincidence  background count rate of 3725 and 105 counts a day respectively for a target mass of 0.64 tons \cite{2}. For a 3.6 ton detector the uncorrelated daily background rate is assumed to be 21,000 counts. This uncorrelated background may be subtracted, thereby contributing a ~150 count uncertainty on the signal.
Due to significant improvements in background rejection from the muon veto system, larger fiducial volume, and inter-event timing cuts the correlated background is found to contribute an additional 200 counts per day \cite{classen}. It is assumed these counts are evenly distributed within 500 keV energy bins.
While the backgrounds cannot be precisely known without an actual deployment, these estimates represent a reasonable extrapolation based on the measured properties of the 3.6 ton detector, and the known signal and background for the SONGS detector.  Other experiments have shown similar signal to background ratios \cite{bulaevskaya2011detection}.

\begin{table}
\protect\caption{Summary of detector parameters used in this analysis. \label{tab:detstuff}}
\begin{ruledtabular}
\begin{tabular}{lccc}
Property & Quantity\\
\hline 
Target Mass & 3.6 Tons \\
Efficiency & 39.2 $\pm 4.7$\% \\
Energy Resolution & $\frac{{20}\%}{\sqrt{E/(MeV)}} $ \\
Fiducial Volume & 100$\%$ \\
Standoff & 25 Meters \\
Overburden & 25 M.W.E \\
Signal to Background & 1262/200 \\
\end{tabular}
\end{ruledtabular}
\end{table}

\begin{figure}
\centering{}\includegraphics[scale=0.45]{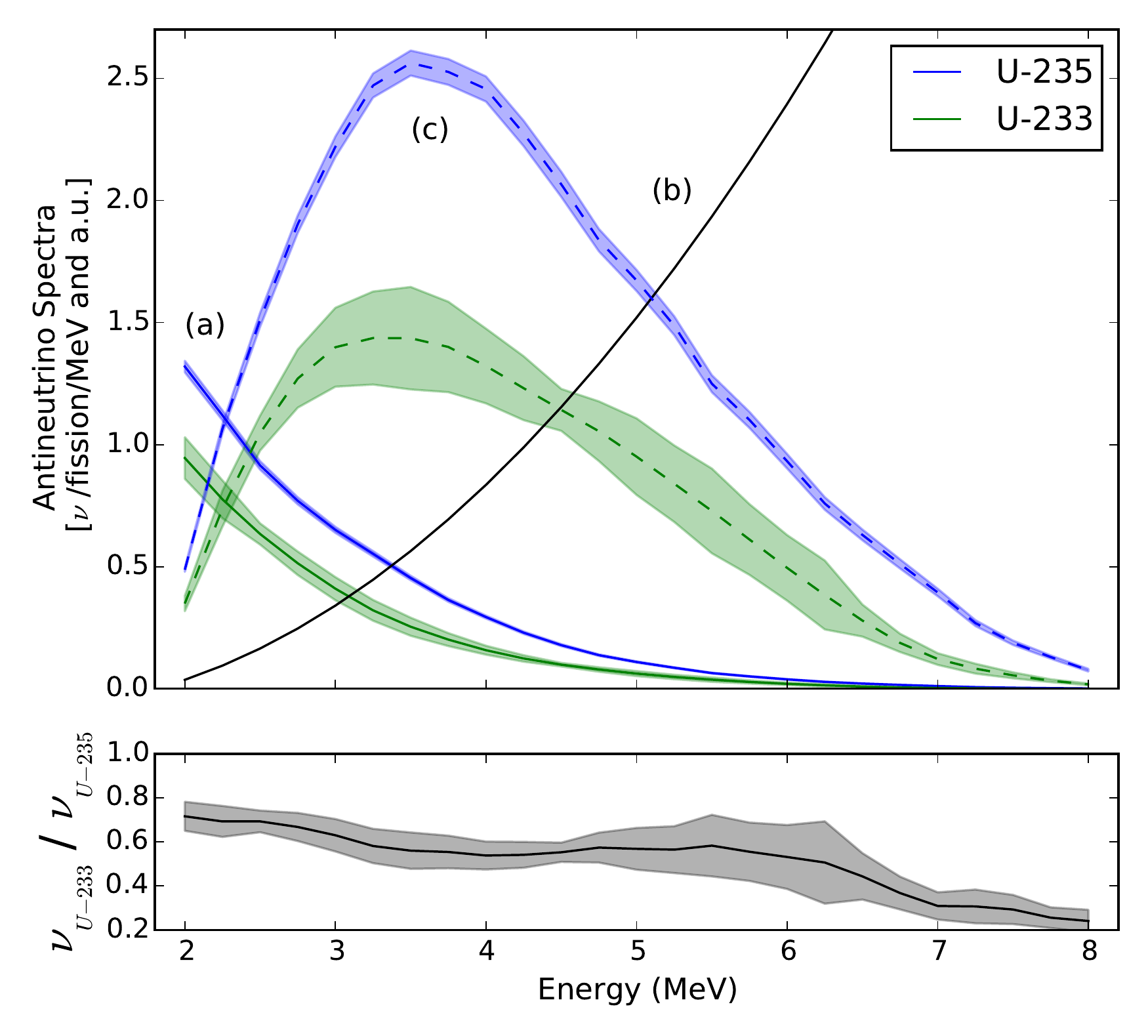}\protect\caption{(Color online).(a) Visualization of antineutrino flux, (b)inverse beta decay cross section, and (c) detected antineutrino events  from \protect\textsuperscript{235}U
and \protect\textsuperscript{233}U fissions (top) shown with 1$\sigma$. Units for the inverse
beta decay cross section and the detected antineutrino events are in
arbitrary units to effectively communicate the effects of physical properties
on the detected spectrum. Oscillation effects are neglected. Ratio of the \protect\textsuperscript{233}U and
\protect\textsuperscript{235}U antineutrino spectra are shown (bottom).
A detailed calculation of the \protect\textsuperscript{233}U antineutrino spectrum
shown above is discussed in Section \ref{sec:Antineutrino-Emission}. 
\label{fig:Visualization-of-antineutrino}}
\end{figure}

Although the antineutrino flux from the reactor is about 10\textsuperscript{21}
$\overline{\nu_{e}}$/s, the low interaction cross section for this
reaction ($\sigma=9.52E^{2}\times10^{-44}cm^{2})$ significantly reduces
the number of events detected \cite{3}. Figure \ref{fig:Visualization-of-antineutrino}
gives a visual representation of the interaction cross section and
antineutrino spectra effect on antineutrino detection for \textsuperscript{235}U and \textsuperscript{233}U
fissions, as well as the ratio between the two. The falling antineutrino spectrum coupled with the quadratic
increase in the interaction cross section results in a unique detection
spectrum for each fissionable isotope \cite{4}. The sum of their
antineutrino spectra can be used to estimate the inventory of fissionable
content in a given reactor throughout the cycle. 

\subsection{Thorium Reactors}

A thorium MSR core consists of a seed and blanket separated by a graphite
moderator. Neutrons emitted from fissile material contained in the
seed, converts \textsuperscript{232}Th in the blanket to \textsuperscript{233}U.
At the beginning of the reactor lifetime the seed is comprised of
low enriched uranium (20\% \textsuperscript{235}U) ; over time enough \textsuperscript{233}U is bred
through the blanket to be used in the seed. \textsuperscript{233}U
is formed from neutron capture by \textsuperscript{232}Th and subsequent
beta decay chains shown below:
\begin{equation}
^{232}Th+n\rightarrow^{233}Th\,\xrightarrow[22.3min]{\beta-}\,^{233}Pa\,\xrightarrow[26.97days]{\beta-}\,^{233}U
\end{equation}
Online reprocessing occurs in the blanket to separate \textsuperscript{233}Pa
and \textsuperscript{233}U from \textsuperscript{232}Th. As shown
in Figure 2, salt from the blanket undergoes flourination to re-inject
uranium into the core, and minimize reactivity losses. \textsuperscript{233}Pa
is then extracted from the salt, and allowed to decay in a sub-critical
storage containment. Another reduction process is used to extract
\textsuperscript{232}Th and replace it in the blanket. After \textsuperscript{233}Pa
decays to \textsuperscript{233}U, it is reintroduced into the core
\cite{6}. Because the conversion ratio for \textsuperscript{232}Th  and \textsuperscript{233}U is
less than one for this reactor, enriched uranium is also added to the seed as makeup fuel. If a state diverts \textsuperscript{233}Pa before its daughter product can be reintroduced into the core, enough \textsuperscript{233}U could be proliferated to construct a nuclear device.

\begin{figure}
\centering{}\includegraphics[scale=0.25]{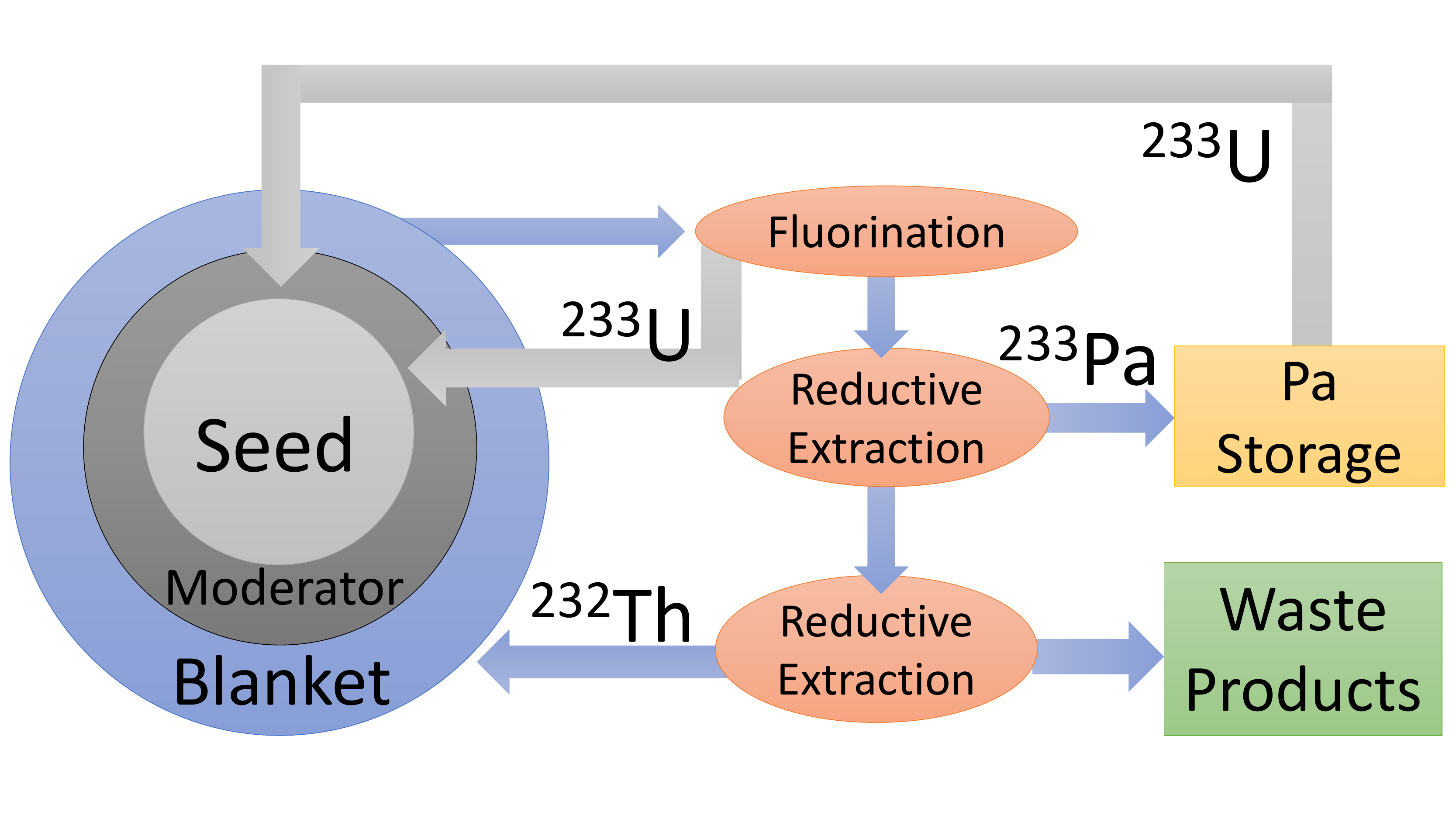}\protect\caption{(Color online). Flow diagram of a thorium MSR reprocessing unit showing the movement
of fissile and fertile material. The introduction of thorium and enriched
uranium come from external processes.}
\end{figure}
 
The reactor used in this analysis is a 500 MWth thermal
thorium MSR taken from the Sandia National Laboratories Nuclear Fuel Cycle Catalog \cite{doe}.
The fuel evolution was calculated using ORIGEN 2.2, a reactor neutronics software to simulate
the isotopic inventory of nuclear systems \cite{origen}.
The fissile inventory of the core over the reactor's lifetime is shown
in Figure 3. The inability to fully refuel the core and the material
degradation from using molten salts reduces the lifetime of these
reactors to a few decades.

\begin{figure}
\includegraphics[scale=0.6]{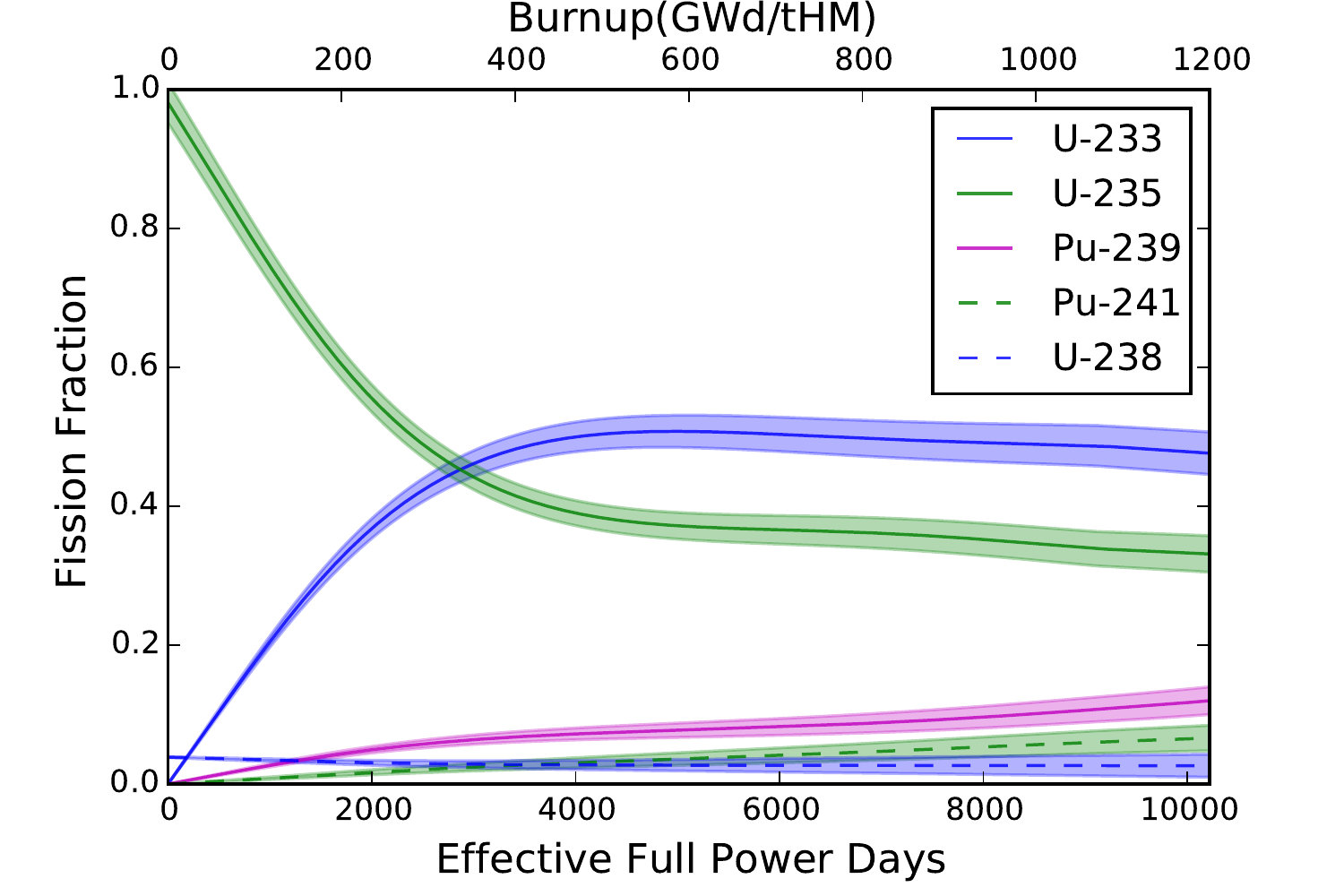}\protect\caption{(Color online). Fission fraction of fissile isotopes throughout the reactor's lifetime with associated uncertanties from simulation. 
At the beginning of the reactor's lifetime, most of the fissions comes
from \protect\textsuperscript{235}U. Over time the contribution of \protect\textsuperscript{233}U
to the fission fraction increases until it reaches an equilibrium.}
\end{figure}

\section{Antineutrino Emission \label{sec:Antineutrino-Emission}}

To determine the expected antineutrino emission from a thorium MSR,
the antineutrino spectra for \textsuperscript{233,235,238}U and \textsuperscript{239,241}Pu
must be evaluated. Because of the graphite moderator the fast neutron
flux in the blanket is negligible. As a result, fissile actinides such as \textsuperscript{232}Th and \textsuperscript{233}Pa do not contribute
to the antineutrino emission from the reactor. The antineutrino spectra
from \textsuperscript{235, 238}U and \textsuperscript{239, 241}Pu have been previously determined by converting the measured electron emission following neutron irradiation at ILL and FRM II \cite{9,u238}; while the antineutrino spectra from the fission of \textsuperscript{233}U is not available in open literature, and must be estimated through an aggregate summation method.

\begin{figure*}
\includegraphics[scale=0.6]{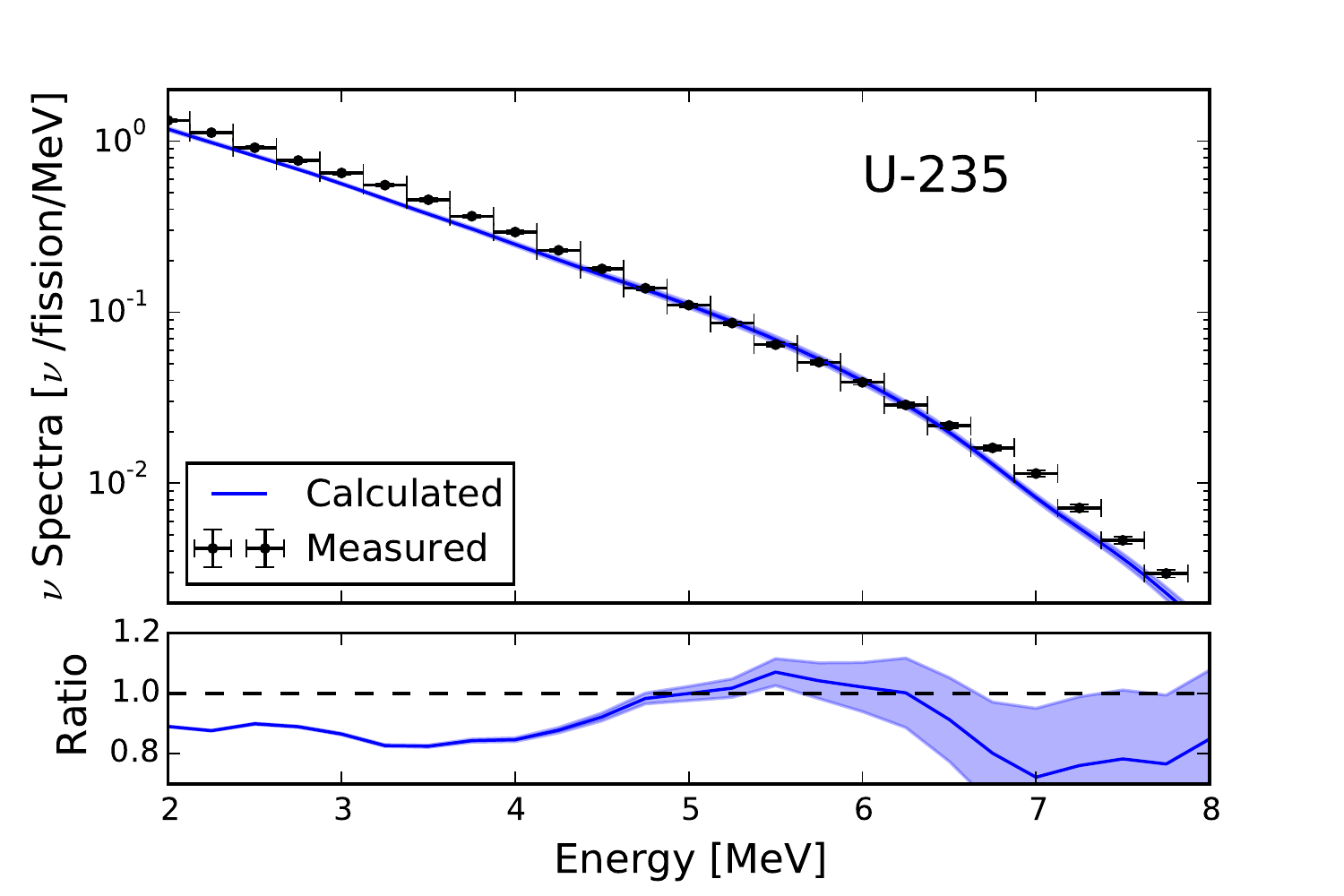}\includegraphics[scale=0.6]{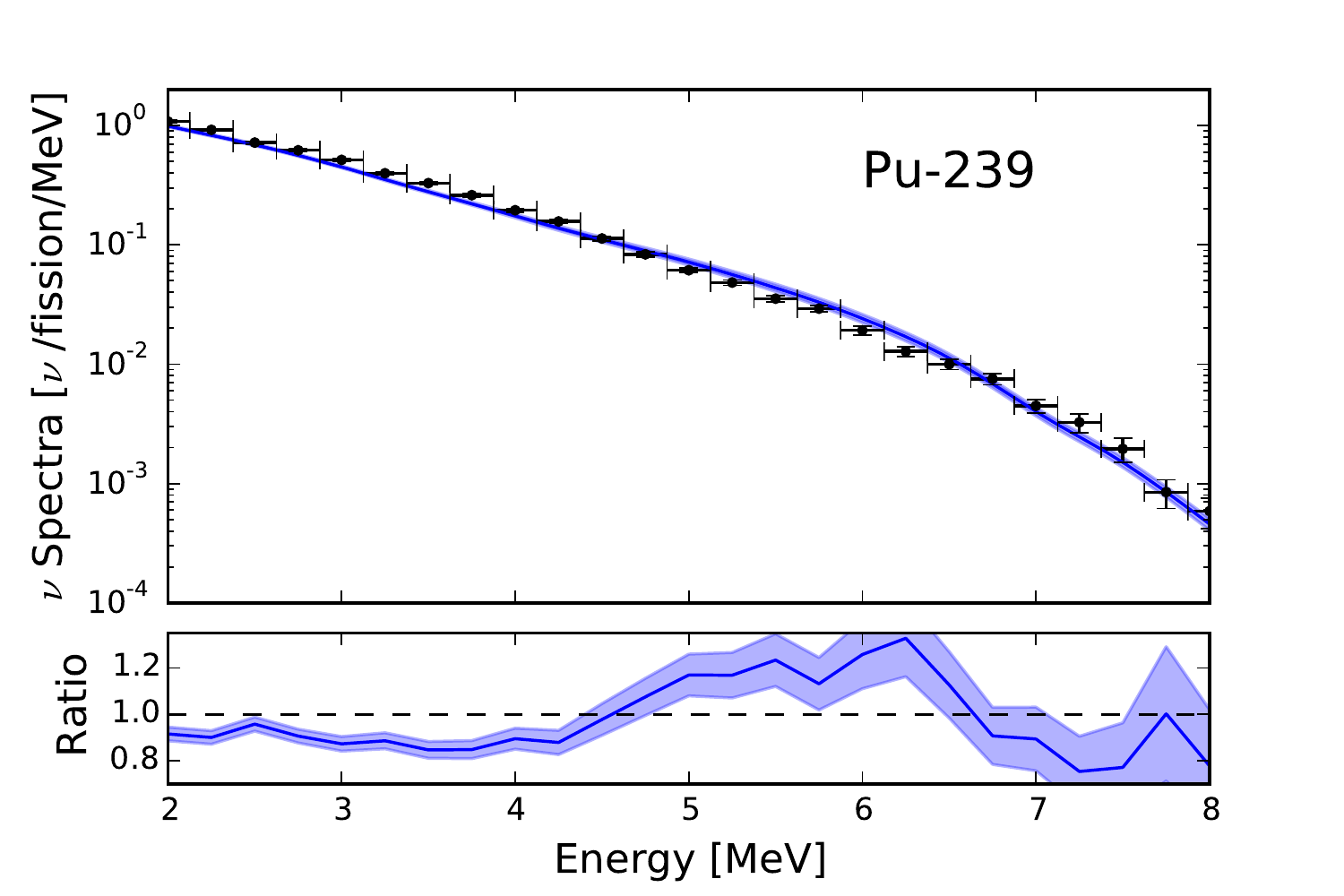}
\includegraphics[scale=0.6]{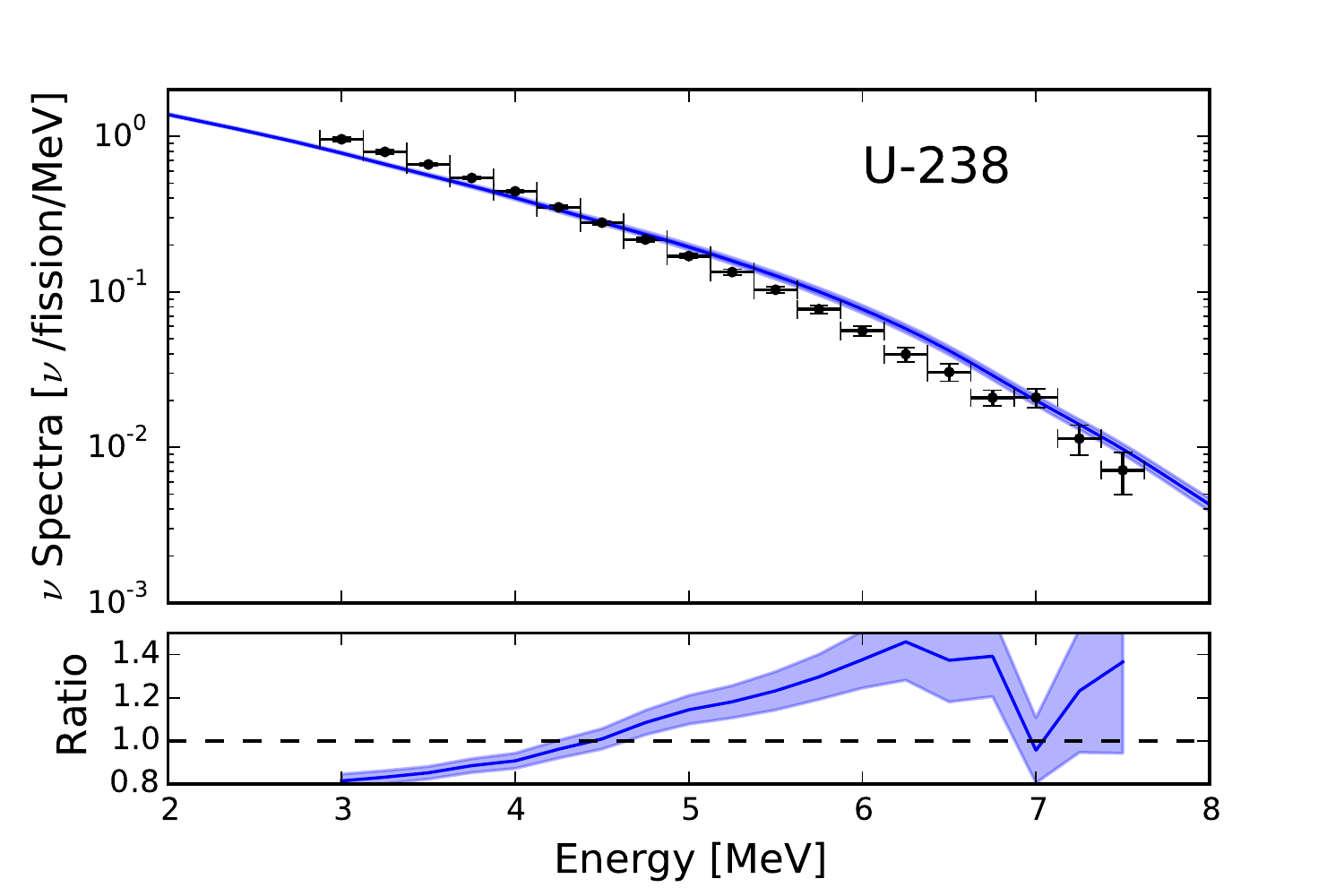}\includegraphics[scale=0.6]{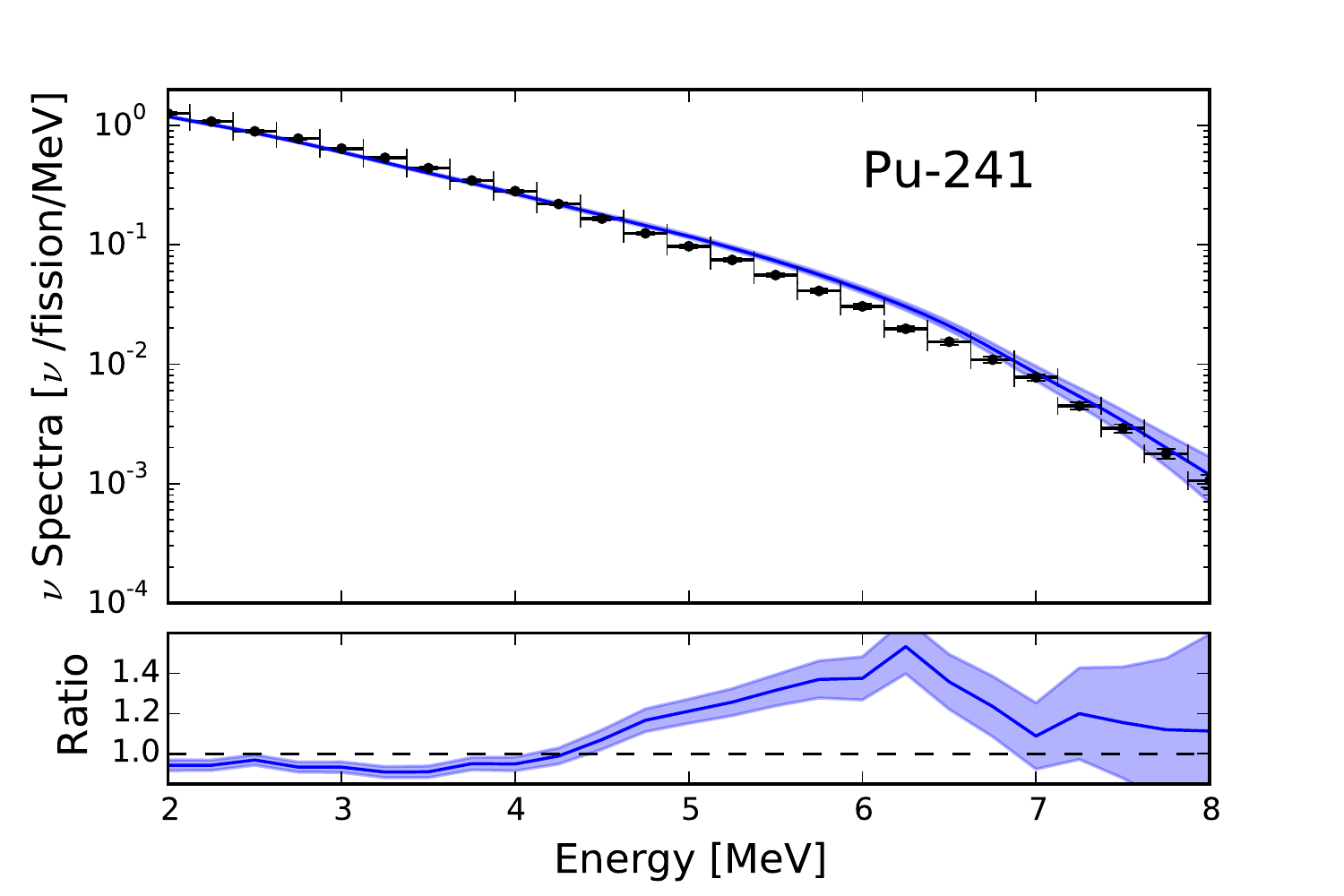}
\protect\caption{(Color online). Comparison of the antineutrino spectra
from the fission of  \textsuperscript{235, 238}U and \textsuperscript{239, 241}Pu determined from 
the summation method (blue) and the converted beta measurements at ILL and FRM II (black). The antineutrino spectra for \textsuperscript{238}U was only evaluated between 3-7.5 MeV \cite{9,u238}  }
\end{figure*}

The antineutrino spectrum from the fission of an isotope, $N(E_{\overline{v}})$
,is an combination of spectra, $P_{\overline{v}}(E_{\overline{v}},E_{0}^{i},Z)$
, from its fission products. In this work, the fission yield and uncertainties was taken from ENDF-349 \cite{england}.
\begin{equation}
N(E_{\overline{v}})=\sum_{n}Y_{n}(Z,A,t)\:\times\sum_{i}b_{n,i}(E_{0}^{i})P_{\overline{v}}(E_{\overline{v}},E_{0}^{i},Z)
\end{equation}
Where $Y_{n}(Z,A,t)$ is the number of beta decays per second for
a given isotope $(Z,A)$; $b_{n,i}(E_{0}^{i})$ is the branching ratio
to an excited state with the electron energy spectrum endpoint: 
\begin{equation}
E_{0}^{i}=Q_{n}-E_{ex}^{i}
\end{equation}
Here $Q_{n} $is the $Q$ value for the beta decay of isotope $(Z,A)$,
and $E_{ex}^{i}$is the excitation energy in the daughter nucleus
$(Z+1,A)$. The decay information for all the fission products was taken from
the Evaluated Nuclear Structure Data Files \cite{NNDC}. The fission products for 
which the decay data was absent were neglected because they only represented less than 10\textsuperscript{-6} of the 
cumulative fission yield, or had such large Q values that these nuclei would be beta delayed neutron candidates such that the antineutrinos emitted from their decay would be less than 3 MeV (below the energy range of this assessment). 

The spectrum shape factor, $P_{\overline{v}}(E_{\overline{v}},E_{0},Z)$
shown in Equation 5 is derived from the phase space relationship between
the electron and antineutrino and the normalization constant, $k$ ,
for each branch; as well as the Fermi Coulomb function, $F(E_{0},Z)$, which
accounts for the Coulomb attraction between the emitted electron and
daughter nucleus is shown in Equation 6 \cite{10}. An allowed Gamow-Teller spectral shape was assumed for all beta 
decays.
\begin{equation}
P_{\overline{v}}  =k\, E_{\overline{v}}^{2}(E_{0}-E_{\overline{v}})^{2}F(E_{0},Z)
\end{equation}

Additional corrections to the antineutrino spectra include both
the electromagnetic and weak-interaction finite size correction, the radiative corrections,
and the weak magnetism correction \cite{finite,4,rad,9}. Effects of the screening correction, were neglected 
do to its small contribution to the antineutrino shape \cite{9}. 
Previous assessments show multiple methods of evaluating the spectrum introduces disagreements in the antineutrino spectrum from fission \cite{hayes}.
To quantify the systematic uncertainty in the aggregate spectra, the antineutrino spectra from \textsuperscript{235, 238}U and \textsuperscript{239, 241}Pu were calculated in the same manner and compared to the ILL and FRM II results \cite{9,u238}. A comparison of the antineutrino spectra from the experiments
at ILL and FRM II compared to those determined using the summation method is shown in Figure 4. The spectra evaluated in this work show a ~10\% deficit from 2-4 MeV, and an excess between 4-7 MeV. The resulting \textsuperscript{233}U antineutrino spectrum is shown in Figure \ref{fig:Visualization-of-antineutrino} and presented in Table \ref{tab:u233t}.

\begin{table}
\protect\caption{Approximated \textsuperscript{233}U antineutrino spectra, and correlated uncertainties using the summation method. \label{tab:u233t}}
\begin{ruledtabular}
\begin{tabular}{lcccc}
$E_{\overline{v}}$ & $N_{\overline{v}}$  & $\sigma$ \\

[MeV] & [$\nu$ fission\textsuperscript{-1} MeV\textsuperscript{-1}] & [\%]\\
\hline 
2.00    &  $9.45\times 10^{-1}$ &  $9.03$ \\
2.25    &  $7.76\times 10^{-1}$ &  $10.0$ \\
2.50    &  $6.34\times 10^{-1}$ &  $6.85$ \\
2.75    &  $5.14\times 10^{-1}$ &  $9.39$ \\
3.00    &  $4.10\times 10^{-1}$ &  $11.6$ \\
3.25    &  $3.22\times 10^{-1}$ &  $13.3$ \\
3.50    &  $2.54\times 10^{-1}$ &  $14.6$ \\
3.75    &  $2.20\times 10^{-1}$ &  $13.2$ \\
4.00    &  $1.58\times 10^{-1}$ &  $11.6$ \\
4.25    &  $1.25\times 10^{-1}$ &  $10.6$ \\
4.50    &  $9.90\times 10^{-2}$ &  $7.52$ \\
4.75    &  $7.92\times 10^{-2}$ &  $12.5$ \\
5.00    &  $6.25\times 10^{-2}$ &  $16.5$ \\
5.25    &  $4.88\times 10^{-2}$ &  $18.7$ \\
5.50    &  $3.77\times 10^{-2}$ &  $23.8$ \\
5.75    &  $2.83\times 10^{-2}$ &  $23.6$ \\
6.00    &  $2.07\times 10^{-2}$ &  $27.1$ \\
6.25    &  $1.45\times 10^{-2}$ &  $36.7$ \\
6.50    &  $9.62\times 10^{-3}$ &  $23.5$ \\
6.75    &  $5.92\times 10^{-3}$ &  $19.9$ \\
7.00    &  $3.53\times 10^{-3}$ &  $19.6$ \\
7.25    &  $2.20\times 10^{-3}$ &  $24.5$ \\
7.50    &  $1.36\times 10^{-3}$ &  $22.1$ \\
7.75    &  $7.61\times 10^{-4}$ &  $17.6$ \\
8.00    &  $3.91\times 10^{-4}$ &  $19.9$ \\
\end{tabular}
\end{ruledtabular}
\end{table}

\section{Diversion Scenario}

Once the antineutrino spectra of fissile isotopes were calculated,
the corresponding detection rate, $D(E_{\overline{v}})$, was found
using Equation \ref{eq:detection}. Here $\rho_{p}$ is the proton
density, $V$ is the detection volume, $\epsilon$ is the detector
efficiency, $T$ is the counting time, $r$ is the distance from
the detector to the reactor core, and $\phi$ is the antineutrino
rate derived from the fission rate in the reactor \citep{2}. The
survival probability, $P_{ee}$, which accounts for the oscillation
of neutrino flavor is also accounted for. However, because the detector is
located 25 m from the reactor core the loss of antineutrinos due to
oscillations was negligible. 
\begin{equation}
D(E_{\overline{v}})=\frac{T\rho_{p}V\epsilon}{4\pi r^{2}}\sigma(E_{\overline{v}})\phi(E_{\bar{v}})P_{ee}(E_{\overline{v}},r)\label{eq:detection}
\end{equation}
 
Once the antineutrino is emitted there is no
way to determine which fissile isotope was responsible for its emission.
Regardless, the count rate evolution and overall detected antineutrino
energy spectrum will still reflect changes in inventory of fissile material in the
reactor. 

The IAEA set limits of concern for unaccounted nuclear material that
can be used for the construction of a nuclear device. In the case
of \textsuperscript{233}U, the IAEA defines a significant quantity
to be 8 kg with a timeliness detection goal of 30 days. Two cases
are compared in the diversion study: a baseline scenario in which
the reactor is under standard operation, and an anomalous scenario
in which a state is trying to divert \textsuperscript{233}U.
The thorium molten salt reactor used in this analysis reaches a \textsuperscript{233}Pa
production equilibrium of about 330 grams a day. In the anomalous
scenario the reactor has reached this equilibrium production and 330
grams of \textsuperscript{233}U is diverted and replaced with 294.36
grams of \textsuperscript{235}U each day until a significant quantity,
8 kg, is obtained. The addition of \textsuperscript{235}U compensates
for the reactivity and power loss from diversion activity. The evaluation
for the analysis takes place at 4500 effective full power days. For this diversion scenario, we assume a proliferating party intends to minimize the amount of time needed
to divert material in order to maximize the time between when a significant
quantity is extracted and when the anomalous activity can be detected.
In this scenario, t = 0 is when the diversion of 330 grams of \textsuperscript{233}U
begins. This continues for twenty-five days, at which point the IAEA
timeliness detection goal for this material is thirty days \cite{glossary2001edition}.
The antineutrino evolution for both the baseline and anomalous case
will be evaluated for 55 days assuming full power operation or until
there is a 95\% confidence that material was diverted.

\subsection{Spectral Analysis\label{sub:Spectral-Analysis}}

A spectral based analysis was performed
by comparing the detected antineutrino spectra from the baseline and anomalous scenarios defined in the previous section, in order to determine  the counting period required to achieve
a 95\% confidence that a significant quantity of material was diverted
. This analysis observes the individual
contributions of each fissioning isotope in the reactor to the overall
detected antineutrino spectrum through an energy binned maximum likelihood analysis for correlated variables:

\begin{equation}
p=\frac{1}{(2\pi)^{n/2}\sqrt{\text{det}(\Sigma)}} \exp\left( -\frac{1}{2} A
\Sigma^{-1} 
A^T \right)
\end{equation}

\begin{figure}
\begin{centering}
\includegraphics[scale=0.45]{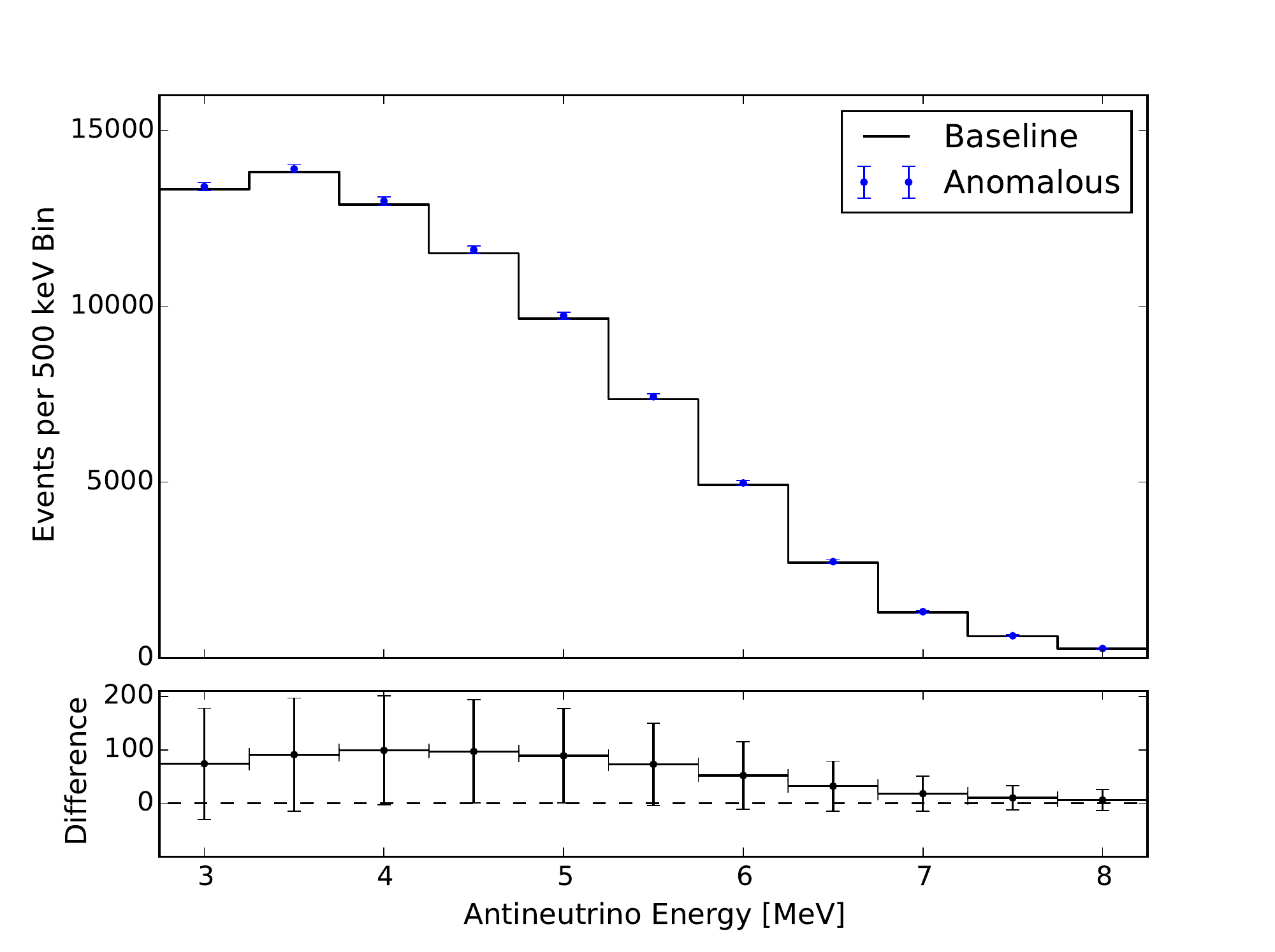}
\par\end{centering}

\protect\caption{(Color online). Detected antineutrino energy spectra integrated for the anomalous and baseline scenario
over an 86 day counting period (top). Energy binned differences between
the anomalous and baseline scenario with statistical errors (bottom).\label{fig:Detected-positron-energy}}
\end{figure}

Here, $A = \left[M_i-B_i,\dots,M_n-B_n \right] $, compares the measured counts, $M$, in each energy bin, $n$, against the counts calculated in the baseline $B$; $\Sigma$ represents the variance-covariance matrix for the measured spectra, and $p$ defines the probability of an anomalous scenario.
Applying the statistics test to the antineutrino spectra shown in Figure \ref{fig:Detected-positron-energy} resulted in an 87\% confidence level for anomalous activity. A counting period of  of 86 days, 31 days past the IAEA timeliness detection goal, was
required to meet the desired confidence level. The ROC, receiver operating characteristic, curve shown in Figure \ref{fig:ROC-curve-for} validates the strength of the test for the spectral analysis. The following section discusses the use of ROC in more detail.

As shown in Fig. \ref{fig:Visualization-of-antineutrino}, the largest spectral difference between \textsuperscript{233}U and \textsuperscript{235}U occur at higher energies where the uncertainties in the spectral shape are the largest. Recent results from the Daya Bay experiment indicate that a small
number of beta-decay isotopes can explain the presence of an anomalous
bump in the antineutrino spectrum around 6-7 MeV \cite{dwyer2015spectral}. In principle, this could have implications to antineutrino detection for nuclear safeguards if the "bump" was attributed to 
specific actinides. However, our results don't depend on the bump and that a higher resolution detector with higher statistics would be needed to exploit this still poorly understood feature. .

Previous studies looking at the diversion of plutonium have shown an excess of counts at higher energies, and a deficit at lower energies due the shape of \textsuperscript{239}Pu $and$ \textsuperscript{241}Pu relative to \textsuperscript{235}U \cite{DPRK}. Because the antineutrino spectrum of \textsuperscript{233}U lower than that of \textsuperscript{235}U at all energies, a spectral analysis shows an excess of counts in each energy bin. Additionally, results from this analysis can be improved if the uncertainties due to counting statistics and the predicted antineutrino emission were reduced.

\subsection{Rate Evolution Analysis\label{sub:Rate-Analysis}}

\begin{figure}
\begin{centering}
\includegraphics[scale=0.6]{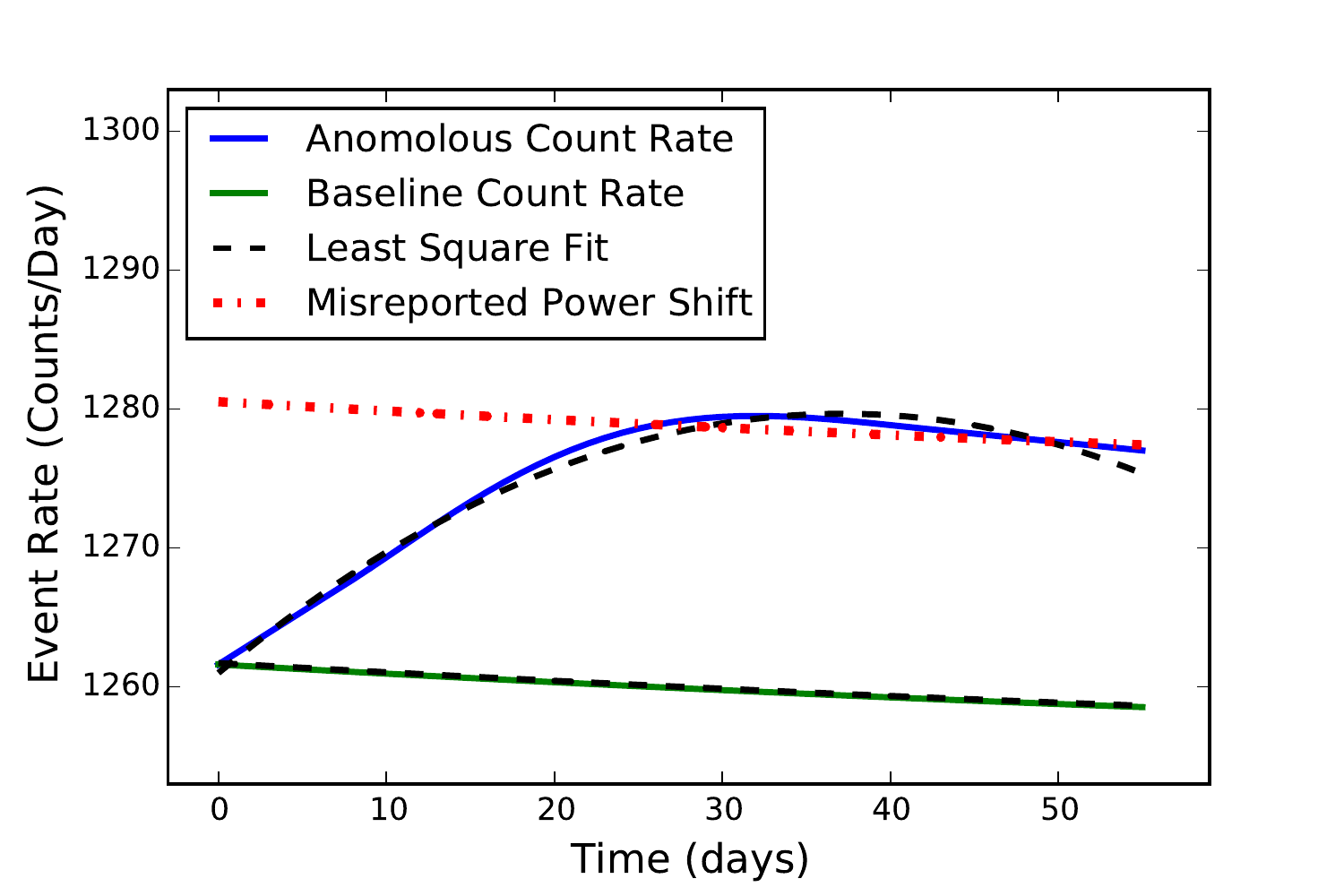}
\par\end{centering}

\protect\caption{(Color online). Baseline (green) and anomalous scenario (blue) evolutions of the antineutrino
count rate in counts per day over the fifty-five day period, along with the expected count rate in the case of operator malice (red).The associated errors in the count rates have
been removed to show a direct comparison of the count rates and calculated fits. \label{fig:Baseline(blue)-and-anomalous} }
\end{figure}

In this analysis the antineutrino count rate evolution in an anomalous scenario
was compared to the expected antineutrino counts under standard operations
(baseline) using a hypothesis testing procedure \cite{bulaevskaya2011detection}. Figure \ref{fig:Baseline(blue)-and-anomalous}
shows the count rate evolution in both the baseline and anomalous
scenarios along with the LS regression fit.
The antineutrino count rate for this thorium MSR shows a smaller decline 
over time than that of the PWR seen in Reference \cite{bulaevskaya2011detection}. 
This MSR is refueled online with more frequency than a PWR, as a result the 
reactivity of a reactor, for a given thermal output, is primarily controlled by the makeup fuel; whereas, in a PWR the boron concentration of the coolant throughout a cycle indiscriminately effects the actinides fission rate. 

Once the antineutrino count rate was determined, a quadratic fit using
least square (LS) regression was applied to the rate. To eliminate
dependencies between the coefficients, the LS regression was performed
on the sample mean $\left(t-\overline{t}\right)$as shown in Equation
\ref{eq:mean ls function}. 
\begin{equation}
N_{\overline{\upsilon}}^{(B)}=\beta_{0}^{(B)}+\beta_{1}^{(B)}\left(t-\overline{t}\right)+\beta_{2}^{(B)}\left(t-\overline{t}\right)^{2}\label{eq:mean ls function}
\end{equation}
Here the superscript are consistant with those used in the spectral analysis.
Comparing the coefficients $\beta_{0}$, $\beta_{1}$,
and $\beta_{2}$ between the measured and baseline fit will indicate
if material has been diverted using the following hypothesis test:

\begin{equation}
H_{0}^{i}:\,\beta_{i}^{(M)}=\beta_{i}^{(B)\quad}\:{\textstyle {\textstyle vrs}}\:\quad H_{a}^{i}:\,\beta_{i}^{(M)}\neq\beta_{i}^{(B)}\;\;\quad \label{eq:hypotheisis}
\end{equation}
Equation \ref{eq:hypotheisis} can be determined using the following
test statistics,

\begin{equation}
s_{i}=\frac{\hat{\beta_{i}^{(M)}}-\hat{\beta_{i}^{(B)}}}{\sqrt{\sigma^{2}(\hat{\beta_{i}^{(M)}})+\sigma^{2}(\hat{\beta_{i}^{(B)}})}}
\end{equation}
and the corresponding $p$ value:
\begin{equation}
p_{i}=2\cdot P(S\geq\left|s_{i}\right|)
\end{equation}
Here $S$ has a Student's $t$ distribution with $2\cdot(n-3)$ degrees
of freedom, with $n$ being the number of count rate measurements and  $\sigma$ referring to the uncertainties in the daily count rate. Although not included in this work, the systematic uncertainties in the detector response and antineutrino rate may be reduced if previous measurements on this reactor can allow for template matching \cite{bulaevskaya2011detection}. 

When applying the test to the count rate evolution, two of the three coefficients reject the null hypothesis
in favor of the alternative hypothesis. The one coefficient, $\beta_{0}$, that showed no statistical difference was related to the absolute antineutrino rate emission. This shows that the test is not dependent on precise knowledge of the expected count rate, but observes general trends (i.e. is the count rate monotonically decreasing or not) in the rate evolution.

\begin{figure}
\begin{centering}
\includegraphics[scale=0.45]{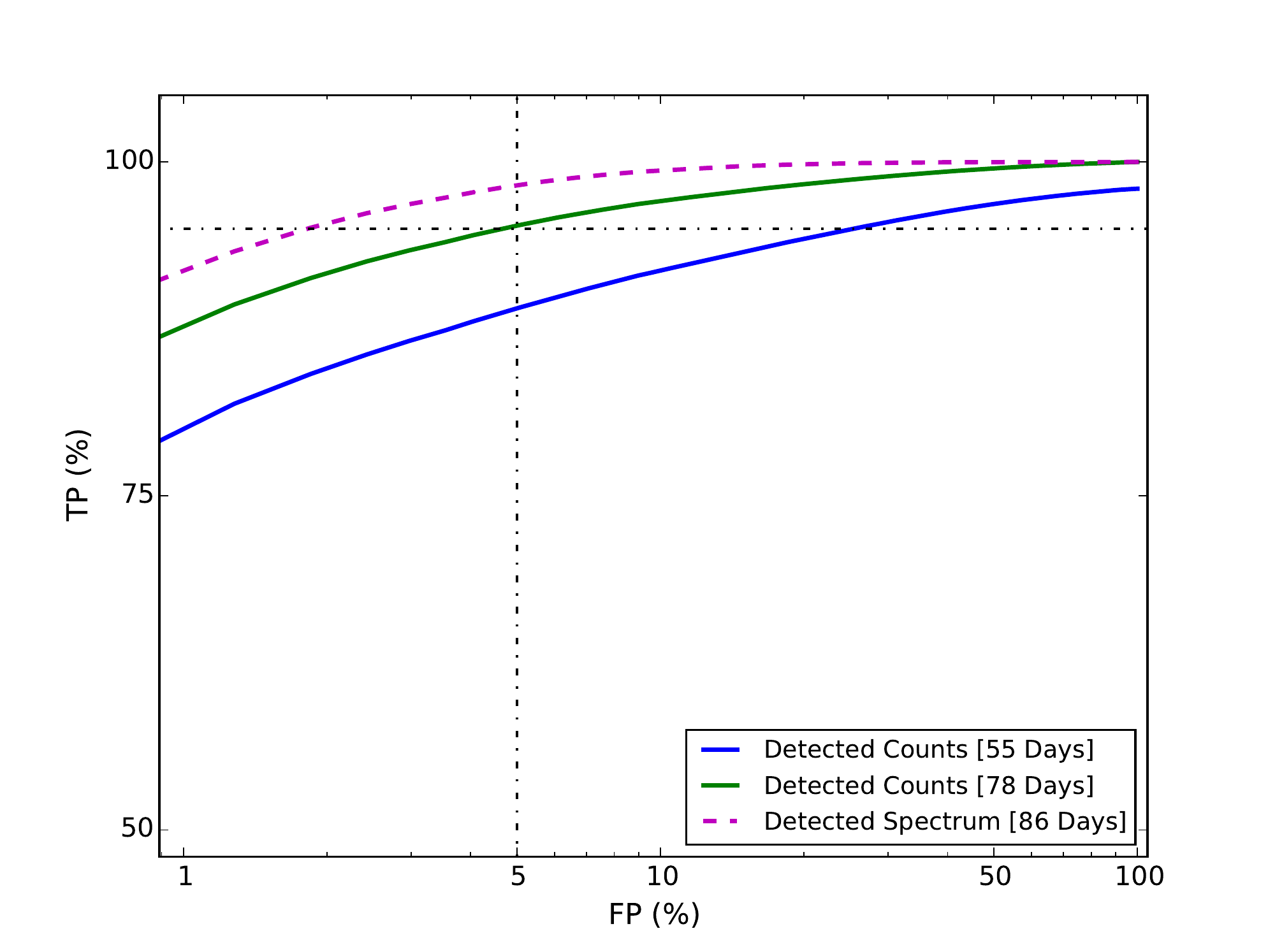}
\par\end{centering}

\protect\caption{(Color online). ROC curve for the  count rate evolution within the IAEA timeliness goal (blue) and when the target TP/FP rate is met for both the count rate evolution (green) and the spectral (purple).  curve for   The desired true positive and false positive rate is plotted with dotted lines.\label{fig:ROC-curve-for} }

\end{figure}

To determine the robustness of the hypothesis tests to statistical variations in LS coefficients,
a Monte Carlo simulation was used to generate one hundred thousand
anomalous and baseline detected antineutrino count rate evolutions
assuming a Gaussian distribution. An ideal threshold
for the true positive/false positive rate was identified as 95\%/5\% . A
true positive result was identified as the test correctly identifying
an anomalous scenario, while a false positive result indicated that
the baseline was incorrectly identified as an anomalous scenario. Figure \ref{fig:ROC-curve-for} shows the ROC curve for the rate analysis under 
a 55 day counting period as well as the counting time required to reach the desired TP/FP rate. Under the 55 day constraint, an 81$\%$/5$\%$ TP/FP rate is achieved. The desired rate is acheived by increasing the counting time to 78 days.  

When this procedure was applied to pressurized water reactors to observe
the diversion of plutonium, it could not identify anomalous activity
as quickly if the operator misreported the thermal power output \cite{bulaevskaya2011detection}. Because
of the transient growth of the antineutrino spectra caused by the diversion of
\textsuperscript{233}U, misreporting the reactor power would not mask proliferation. Figure \ref{fig:Baseline(blue)-and-anomalous} shows the expected antineutrino count rate 
if the reactor was operating at a 7MW excess. Here the
 the $\beta_{0}$ coefficients are matched, requiring more time to detect material diversion.
 After applying the same procedure, as mentioned above, an anomalous scenario can be detected within 114 days in the presence of a false power report.
Additionally, the effects of misreported power shifts can give insight to the test's sensitivity to detector drifts. The detector used in this analysis has not been deployed long enough 
to study these effects, but the SONGS1 detector showed a less than 1\% drift while taking data \cite{drift}.

\begin{table}
\protect\caption{Sensitivity of the count rate evolution test using Monte Carlo simulations. The results show the false positive rate that would incur for a 5\% false positive rate for both the diversion scenario presented in the study, and a misreported power shift.\label{tab:drift}}
\begin{ruledtabular}
\begin{tabular}{ccccc}
Counting & Diversion Scenario & Misreported Power\\
Time   & TP/FP & TP/FP \\
\hline 
55 Days & 81\%/5\% & 58\%/5\% \\
78 Days& 95\% /5\% & 73\% /5\% \\
114 Days & 99\% /2\% & 95\%/5\% \\
\end{tabular}
\end{ruledtabular}
\end{table}

\section{Conclusion}

The antineutrino emission from a thorium MSR was analyzed to determine
if the diversion of a significant quantity of \textsuperscript{233}U
could be observed within the IAEA timeliness goal. In order to perform the analysis, we calculated the antineutrino emission spectrum of \textsuperscript{233}U, based on the fission product yields. Using a spectral
and rate-time based analysis the diversion \textsuperscript{233}U could
be detected 61 and 53 days after the diversion of 8
kg of \textsuperscript{233}U respectively; in a total counting period of 86 and 78 days. 
Usually a spectral analysis is
more sensitive to diversion than the rate-time analysis because it provides
more information about the antineutrino emission. In this analysis
the integral spectrum over the entire 55 day period is being binned
into a single histogram so the day to day changes in the rate of each
bin is averaged out.

Evaluating the antineutrino count rate gives insight to the power
level of the reactor, where analyzing the detected antineutrino spectra
is a reflection of its isotopic concentration. Although this method
requires less time, relying solely on the count rate makes the analysis
susceptible to operator malfeasance. When using the antineutrino count
rate, there is a trade off between exceeding IAEA timeliness
goals and being sensitive to false declarations of reactor power.

The test for material diversion was heavily dependent on counting
statistics. Desirable test performances are attainable through feasible
improvements to antineutrino detectors, such as better instrumentation
for small scale antineutrino detectors or increasing the detector
mass. Addtionally, variations in thorium reactor designs may also effect the sensitivity to diversion.
Some models incorporate scheduled refueling, reducing the counting
time from 55 days to 30 days; or utilize epithermal and fast neutron
energies, which would effect the antineutrino emission from fission.
Other designs incorporating solid fuels do not separate the \textsuperscript{233}Pa from the blanket. 
This changes the SNM from unirradiated direct use material to irradiated use material, allowing 
for longer diversion detection constraints.

\begin{acknowledgments}
This work was performed under the auspices of the U.S.
Department of Energy by Lawrence Livermore National Laboratory
under contract DE-AC52-07NA27344. LLNL-JRNL-646478, and 
supported by the Department of Energy National Nuclear Security Administration under Award Number DE-NA0000979.
\end{acknowledgments}


\bibliography{AMTR}

\begin{thebibliography}{21}%
\makeatletter
\providecommand \@ifxundefined [1]{%
 \@ifx{#1\undefined}
}%
\providecommand \@ifnum [1]{%
 \ifnum #1\expandafter \@firstoftwo
 \else \expandafter \@secondoftwo
 \fi
}%
\providecommand \@ifx [1]{%
 \ifx #1\expandafter \@firstoftwo
 \else \expandafter \@secondoftwo
 \fi
}%
\providecommand \natexlab [1]{#1}%
\providecommand \enquote  [1]{``#1''}%
\providecommand \bibnamefont  [1]{#1}%
\providecommand \bibfnamefont [1]{#1}%
\providecommand \citenamefont [1]{#1}%
\providecommand \href@noop [0]{\@secondoftwo}%
\providecommand \href [0]{\begingroup \@sanitize@url \@href}%
\providecommand \@href[1]{\@@startlink{#1}\@@href}%
\providecommand \@@href[1]{\endgroup#1\@@endlink}%
\providecommand \@sanitize@url [0]{\catcode `\\12\catcode `\$12\catcode
  `\&12\catcode `\#12\catcode `\^12\catcode `\_12\catcode `\%12\relax}%
\providecommand \@@startlink[1]{}%
\providecommand \@@endlink[0]{}%
\providecommand \url  [0]{\begingroup\@sanitize@url \@url }%
\providecommand \@url [1]{\endgroup\@href {#1}{\urlprefix }}%
\providecommand \urlprefix  [0]{URL }%
\providecommand \Eprint [0]{\href }%
\providecommand \doibase [0]{http://dx.doi.org/}%
\providecommand \selectlanguage [0]{\@gobble}%
\providecommand \bibinfo  [0]{\@secondoftwo}%
\providecommand \bibfield  [0]{\@secondoftwo}%
\providecommand \translation [1]{[#1]}%
\providecommand \BibitemOpen [0]{}%
\providecommand \bibitemStop [0]{}%
\providecommand \bibitemNoStop [0]{.\EOS\space}%
\providecommand \EOS [0]{\spacefactor3000\relax}%
\providecommand \BibitemShut  [1]{\csname bibitem#1\endcsname}%
\let\auto@bib@innerbib\@empty
\bibitem [{1(2014)}]{1}%
  \BibitemOpen
  \href@noop {} {\bibfield  {journal} {\bibinfo  {journal} {IAEA Safety Report
  Series}\ } (\bibinfo {year} {2014})}\BibitemShut {NoStop}%
\bibitem [{\citenamefont {Bowden}\ \emph {et~al.}(2007)\citenamefont {Bowden},
  \citenamefont {Bernstein}, \citenamefont {Allen}, \citenamefont {Brennan},
  \citenamefont {Cunningham}, \citenamefont {Estrada}, \citenamefont {Greaves},
  \citenamefont {Hagmann}, \citenamefont {Lund}, \citenamefont {Mengesha} \emph
  {et~al.}}]{2}%
  \BibitemOpen
  \bibfield  {author} {\bibinfo {author} {\bibfnamefont {N.}~\bibnamefont
  {Bowden}}, \bibinfo {author} {\bibfnamefont {A.}~\bibnamefont {Bernstein}},
  \bibinfo {author} {\bibfnamefont {M.}~\bibnamefont {Allen}}, \bibinfo
  {author} {\bibfnamefont {J.}~\bibnamefont {Brennan}}, \bibinfo {author}
  {\bibfnamefont {M.}~\bibnamefont {Cunningham}}, \bibinfo {author}
  {\bibfnamefont {J.}~\bibnamefont {Estrada}}, \bibinfo {author} {\bibfnamefont
  {C.}~\bibnamefont {Greaves}}, \bibinfo {author} {\bibfnamefont
  {C.}~\bibnamefont {Hagmann}}, \bibinfo {author} {\bibfnamefont
  {J.}~\bibnamefont {Lund}}, \bibinfo {author} {\bibfnamefont {W.}~\bibnamefont
  {Mengesha}},  \emph {et~al.},\ }\href@noop {} {\bibfield  {journal} {\bibinfo
   {journal} {Nuclear Instruments and Methods in Physics Research Section A:
  Accelerators, Spectrometers, Detectors and Associated Equipment}\ }\textbf
  {\bibinfo {volume} {572}},\ \bibinfo {pages} {985} (\bibinfo {year}
  {2007})}\BibitemShut {NoStop}%
\bibitem [{\citenamefont {Classen}\ \emph {et~al.}(2015)\citenamefont
  {Classen}, \citenamefont {Bernstein}, \citenamefont {Bowden}, \citenamefont
  {Cabrera-Palmer}, \citenamefont {Ho}, \citenamefont {Jonkmans}, \citenamefont
  {Kogler}, \citenamefont {Reyna},\ and\ \citenamefont {Sur}}]{classen}%
  \BibitemOpen
  \bibfield  {author} {\bibinfo {author} {\bibfnamefont {T.}~\bibnamefont
  {Classen}}, \bibinfo {author} {\bibfnamefont {A.}~\bibnamefont {Bernstein}},
  \bibinfo {author} {\bibfnamefont {N.}~\bibnamefont {Bowden}}, \bibinfo
  {author} {\bibfnamefont {B.}~\bibnamefont {Cabrera-Palmer}}, \bibinfo
  {author} {\bibfnamefont {A.}~\bibnamefont {Ho}}, \bibinfo {author}
  {\bibfnamefont {G.}~\bibnamefont {Jonkmans}}, \bibinfo {author}
  {\bibfnamefont {L.}~\bibnamefont {Kogler}}, \bibinfo {author} {\bibfnamefont
  {D.}~\bibnamefont {Reyna}}, \ and\ \bibinfo {author} {\bibfnamefont
  {B.}~\bibnamefont {Sur}},\ }\href@noop {} {\bibfield  {journal} {\bibinfo
  {journal} {Nuclear Instruments and Methods in Physics Research Section A:
  Accelerators, Spectrometers, Detectors and Associated Equipment}\ }\textbf
  {\bibinfo {volume} {771}},\ \bibinfo {pages} {139} (\bibinfo {year}
  {2015})}\BibitemShut {NoStop}%
\bibitem [{\citenamefont {Bulaevskaya}\ and\ \citenamefont
  {Bernstein}(2011)}]{bulaevskaya2011detection}%
  \BibitemOpen
  \bibfield  {author} {\bibinfo {author} {\bibfnamefont {V.}~\bibnamefont
  {Bulaevskaya}}\ and\ \bibinfo {author} {\bibfnamefont {A.}~\bibnamefont
  {Bernstein}},\ }\href@noop {} {\bibfield  {journal} {\bibinfo  {journal}
  {Journal of Applied Physics}\ }\textbf {\bibinfo {volume} {109}},\ \bibinfo
  {pages} {114909} (\bibinfo {year} {2011})}\BibitemShut {NoStop}%
\bibitem [{\citenamefont {Bemporad}\ \emph {et~al.}(2002)\citenamefont
  {Bemporad}, \citenamefont {Gratta},\ and\ \citenamefont {Vogel}}]{3}%
  \BibitemOpen
  \bibfield  {author} {\bibinfo {author} {\bibfnamefont {C.}~\bibnamefont
  {Bemporad}}, \bibinfo {author} {\bibfnamefont {G.}~\bibnamefont {Gratta}}, \
  and\ \bibinfo {author} {\bibfnamefont {P.}~\bibnamefont {Vogel}},\
  }\href@noop {} {\bibfield  {journal} {\bibinfo  {journal} {Reviews of Modern
  Physics}\ }\textbf {\bibinfo {volume} {74}},\ \bibinfo {pages} {297}
  (\bibinfo {year} {2002})}\BibitemShut {NoStop}%
\bibitem [{\citenamefont {Mueller}\ \emph {et~al.}(2011)\citenamefont
  {Mueller}, \citenamefont {Lhuillier}, \citenamefont {Fallot}, \citenamefont
  {Letourneau}, \citenamefont {Cormon}, \citenamefont {Fechner}, \citenamefont
  {Giot}, \citenamefont {Lasserre}, \citenamefont {Martino}, \citenamefont
  {Mention} \emph {et~al.}}]{4}%
  \BibitemOpen
  \bibfield  {author} {\bibinfo {author} {\bibfnamefont {T.}~\bibnamefont
  {Mueller}}, \bibinfo {author} {\bibfnamefont {D.}~\bibnamefont {Lhuillier}},
  \bibinfo {author} {\bibfnamefont {M.}~\bibnamefont {Fallot}}, \bibinfo
  {author} {\bibfnamefont {A.}~\bibnamefont {Letourneau}}, \bibinfo {author}
  {\bibfnamefont {S.}~\bibnamefont {Cormon}}, \bibinfo {author} {\bibfnamefont
  {M.}~\bibnamefont {Fechner}}, \bibinfo {author} {\bibfnamefont
  {L.}~\bibnamefont {Giot}}, \bibinfo {author} {\bibfnamefont {T.}~\bibnamefont
  {Lasserre}}, \bibinfo {author} {\bibfnamefont {J.}~\bibnamefont {Martino}},
  \bibinfo {author} {\bibfnamefont {G.}~\bibnamefont {Mention}},  \emph
  {et~al.},\ }\href@noop {} {\bibfield  {journal} {\bibinfo  {journal}
  {Physical Review C}\ }\textbf {\bibinfo {volume} {83}},\ \bibinfo {pages}
  {054615} (\bibinfo {year} {2011})}\BibitemShut {NoStop}%
\bibitem [{\citenamefont {Nuttin}\ \emph {et~al.}(2005)\citenamefont {Nuttin},
  \citenamefont {Heuer}, \citenamefont {Billebaud}, \citenamefont {Brissot},
  \citenamefont {Le~Brun}, \citenamefont {Liatard}, \citenamefont {Loiseaux},
  \citenamefont {Mathieu}, \citenamefont {Meplan}, \citenamefont
  {Merle-Lucotte} \emph {et~al.}}]{6}%
  \BibitemOpen
  \bibfield  {author} {\bibinfo {author} {\bibfnamefont {A.}~\bibnamefont
  {Nuttin}}, \bibinfo {author} {\bibfnamefont {D.}~\bibnamefont {Heuer}},
  \bibinfo {author} {\bibfnamefont {A.}~\bibnamefont {Billebaud}}, \bibinfo
  {author} {\bibfnamefont {R.}~\bibnamefont {Brissot}}, \bibinfo {author}
  {\bibfnamefont {C.}~\bibnamefont {Le~Brun}}, \bibinfo {author} {\bibfnamefont
  {E.}~\bibnamefont {Liatard}}, \bibinfo {author} {\bibfnamefont
  {J.}~\bibnamefont {Loiseaux}}, \bibinfo {author} {\bibfnamefont
  {L.}~\bibnamefont {Mathieu}}, \bibinfo {author} {\bibfnamefont
  {O.}~\bibnamefont {Meplan}}, \bibinfo {author} {\bibfnamefont
  {E.}~\bibnamefont {Merle-Lucotte}},  \emph {et~al.},\ }\href@noop {}
  {\bibfield  {journal} {\bibinfo  {journal} {Progress in Nuclear Energy}\
  }\textbf {\bibinfo {volume} {46}},\ \bibinfo {pages} {77} (\bibinfo {year}
  {2005})}\BibitemShut {NoStop}%
\bibitem [{\citenamefont {Wigeland}\ \emph {et~al.}(2014)\citenamefont
  {Wigeland}, \citenamefont {Taiwo}, \citenamefont {Ludewig}, \citenamefont
  {Todosow}, \citenamefont {Halsey}, \citenamefont {Gehin}, \citenamefont
  {Jubin}, \citenamefont {Buelt}, \citenamefont {Stockinger}, \citenamefont
  {Jenni},\ and\ \citenamefont {Oakley}}]{doe}%
  \BibitemOpen
  \bibfield  {author} {\bibinfo {author} {\bibfnamefont {R.}~\bibnamefont
  {Wigeland}}, \bibinfo {author} {\bibfnamefont {T.}~\bibnamefont {Taiwo}},
  \bibinfo {author} {\bibfnamefont {H.}~\bibnamefont {Ludewig}}, \bibinfo
  {author} {\bibfnamefont {M.}~\bibnamefont {Todosow}}, \bibinfo {author}
  {\bibfnamefont {W.}~\bibnamefont {Halsey}}, \bibinfo {author} {\bibfnamefont
  {J.}~\bibnamefont {Gehin}}, \bibinfo {author} {\bibfnamefont
  {R.}~\bibnamefont {Jubin}}, \bibinfo {author} {\bibfnamefont
  {J.}~\bibnamefont {Buelt}}, \bibinfo {author} {\bibfnamefont
  {S.}~\bibnamefont {Stockinger}}, \bibinfo {author} {\bibfnamefont
  {K.}~\bibnamefont {Jenni}}, \ and\ \bibinfo {author} {\bibfnamefont
  {B.}~\bibnamefont {Oakley}},\ }\href@noop {} {\  (\bibinfo {year}
  {2014})}\BibitemShut {NoStop}%
\bibitem [{\citenamefont {Gauld}\ \emph {et~al.}(2009)\citenamefont {Gauld},
  \citenamefont {Hermann},\ and\ \citenamefont {Westfall}}]{origen}%
  \BibitemOpen
  \bibfield  {author} {\bibinfo {author} {\bibfnamefont {I.}~\bibnamefont
  {Gauld}}, \bibinfo {author} {\bibfnamefont {O.}~\bibnamefont {Hermann}}, \
  and\ \bibinfo {author} {\bibfnamefont {R.}~\bibnamefont {Westfall}},\
  }\href@noop {} {\bibfield  {journal} {\bibinfo  {journal} {ORNL/TM-2005/39,
  Version}\ }\textbf {\bibinfo {volume} {6}} (\bibinfo {year}
  {2009})}\BibitemShut {NoStop}%
\bibitem [{\citenamefont {Huber}(2011)}]{9}%
  \BibitemOpen
  \bibfield  {author} {\bibinfo {author} {\bibfnamefont {P.}~\bibnamefont
  {Huber}},\ }\href@noop {} {\bibfield  {journal} {\bibinfo  {journal}
  {Physical Review C}\ }\textbf {\bibinfo {volume} {84}},\ \bibinfo {pages}
  {024617} (\bibinfo {year} {2011})}\BibitemShut {NoStop}%
\bibitem [{\citenamefont {Haag}\ \emph {et~al.}(2014)\citenamefont {Haag},
  \citenamefont {G{\"u}tlein}, \citenamefont {Hofmann}, \citenamefont
  {Oberauer}, \citenamefont {Potzel}, \citenamefont {Schreckenbach},\ and\
  \citenamefont {Wagner}}]{u238}%
  \BibitemOpen
  \bibfield  {author} {\bibinfo {author} {\bibfnamefont {N.}~\bibnamefont
  {Haag}}, \bibinfo {author} {\bibfnamefont {A.}~\bibnamefont {G{\"u}tlein}},
  \bibinfo {author} {\bibfnamefont {M.}~\bibnamefont {Hofmann}}, \bibinfo
  {author} {\bibfnamefont {L.}~\bibnamefont {Oberauer}}, \bibinfo {author}
  {\bibfnamefont {W.}~\bibnamefont {Potzel}}, \bibinfo {author} {\bibfnamefont
  {K.}~\bibnamefont {Schreckenbach}}, \ and\ \bibinfo {author} {\bibfnamefont
  {F.}~\bibnamefont {Wagner}},\ }\href@noop {} {\bibfield  {journal} {\bibinfo
  {journal} {Physical review letters}\ }\textbf {\bibinfo {volume} {112}},\
  \bibinfo {pages} {122501} (\bibinfo {year} {2014})}\BibitemShut {NoStop}%
\bibitem [{\citenamefont {England}\ and\ \citenamefont
  {Rider}(1994)}]{england}%
  \BibitemOpen
  \bibfield  {author} {\bibinfo {author} {\bibfnamefont {T.}~\bibnamefont
  {England}}\ and\ \bibinfo {author} {\bibfnamefont {B.}~\bibnamefont
  {Rider}},\ }\href@noop {} {\bibfield  {journal} {\bibinfo  {journal}
  {ENDF-349, LA-UR-94-3106, Los Alamos National Laboratory}\ } (\bibinfo {year}
  {1994})}\BibitemShut {NoStop}%
\bibitem [{\citenamefont {Bhat}(1992)}]{NNDC}%
  \BibitemOpen
  \bibfield  {author} {\bibinfo {author} {\bibfnamefont {M.}~\bibnamefont
  {Bhat}},\ }in\ \href@noop {} {\emph {\bibinfo {booktitle} {Nuclear Data for
  Science and Technology}}}\ (\bibinfo {organization} {Springer},\ \bibinfo
  {year} {1992})\ pp.\ \bibinfo {pages} {817--821}\BibitemShut {NoStop}%
\bibitem [{\citenamefont {Vogel}\ \emph {et~al.}(1981)\citenamefont {Vogel},
  \citenamefont {Schenter}, \citenamefont {Mann},\ and\ \citenamefont
  {Schenter}}]{10}%
  \BibitemOpen
  \bibfield  {author} {\bibinfo {author} {\bibfnamefont {P.}~\bibnamefont
  {Vogel}}, \bibinfo {author} {\bibfnamefont {G.}~\bibnamefont {Schenter}},
  \bibinfo {author} {\bibfnamefont {F.}~\bibnamefont {Mann}}, \ and\ \bibinfo
  {author} {\bibfnamefont {R.}~\bibnamefont {Schenter}},\ }\href@noop {}
  {\bibfield  {journal} {\bibinfo  {journal} {Physical Review C}\ }\textbf
  {\bibinfo {volume} {24}},\ \bibinfo {pages} {1543} (\bibinfo {year}
  {1981})}\BibitemShut {NoStop}%
\bibitem [{\citenamefont {Wilkinson}(1990)}]{finite}%
  \BibitemOpen
  \bibfield  {author} {\bibinfo {author} {\bibfnamefont {D.}~\bibnamefont
  {Wilkinson}},\ }\href@noop {} {\bibfield  {journal} {\bibinfo  {journal}
  {Nuclear Instruments and Methods in Physics Research Section A: Accelerators,
  Spectrometers, Detectors and Associated Equipment}\ }\textbf {\bibinfo
  {volume} {290}},\ \bibinfo {pages} {509} (\bibinfo {year}
  {1990})}\BibitemShut {NoStop}%
\bibitem [{\citenamefont {Sirlin}(2011)}]{rad}%
  \BibitemOpen
  \bibfield  {author} {\bibinfo {author} {\bibfnamefont {A.}~\bibnamefont
  {Sirlin}},\ }\href@noop {} {\bibfield  {journal} {\bibinfo  {journal}
  {Physical Review D}\ }\textbf {\bibinfo {volume} {84}},\ \bibinfo {pages}
  {014021} (\bibinfo {year} {2011})}\BibitemShut {NoStop}%
\bibitem [{\citenamefont {Hayes}\ \emph {et~al.}(2014)\citenamefont {Hayes},
  \citenamefont {Friar}, \citenamefont {Garvey}, \citenamefont {Jungman},\ and\
  \citenamefont {Jonkmans}}]{hayes}%
  \BibitemOpen
  \bibfield  {author} {\bibinfo {author} {\bibfnamefont {A.}~\bibnamefont
  {Hayes}}, \bibinfo {author} {\bibfnamefont {J.}~\bibnamefont {Friar}},
  \bibinfo {author} {\bibfnamefont {G.}~\bibnamefont {Garvey}}, \bibinfo
  {author} {\bibfnamefont {G.}~\bibnamefont {Jungman}}, \ and\ \bibinfo
  {author} {\bibfnamefont {G.}~\bibnamefont {Jonkmans}},\ }\href@noop {}
  {\bibfield  {journal} {\bibinfo  {journal} {Physical Review Letters}\
  }\textbf {\bibinfo {volume} {112}},\ \bibinfo {pages} {202501} (\bibinfo
  {year} {2014})}\BibitemShut {NoStop}%
\bibitem [{\citenamefont {Glossary}(2001)}]{glossary2001edition}%
  \BibitemOpen
  \bibfield  {author} {\bibinfo {author} {\bibfnamefont {I.~S.}\ \bibnamefont
  {Glossary}},\ }\href@noop {} {\bibfield  {journal} {\bibinfo  {journal}
  {IAEA, Austria}\ } (\bibinfo {year} {2001})}\BibitemShut {NoStop}%
\bibitem [{\citenamefont {Dwyer}\ and\ \citenamefont
  {Langford}(2015)}]{dwyer2015spectral}%
  \BibitemOpen
  \bibfield  {author} {\bibinfo {author} {\bibfnamefont {D.}~\bibnamefont
  {Dwyer}}\ and\ \bibinfo {author} {\bibfnamefont {T.}~\bibnamefont
  {Langford}},\ }\href@noop {} {\bibfield  {journal} {\bibinfo  {journal}
  {Physical Review Letters}\ }\textbf {\bibinfo {volume} {114}},\ \bibinfo
  {pages} {012502} (\bibinfo {year} {2015})}\BibitemShut {NoStop}%
\bibitem [{\citenamefont {Christensen}\ \emph {et~al.}(2014)\citenamefont
  {Christensen}, \citenamefont {Huber}, \citenamefont {Jaffke},\ and\
  \citenamefont {Shea}}]{DPRK}%
  \BibitemOpen
  \bibfield  {author} {\bibinfo {author} {\bibfnamefont {E.}~\bibnamefont
  {Christensen}}, \bibinfo {author} {\bibfnamefont {P.}~\bibnamefont {Huber}},
  \bibinfo {author} {\bibfnamefont {P.}~\bibnamefont {Jaffke}}, \ and\ \bibinfo
  {author} {\bibfnamefont {T.}~\bibnamefont {Shea}},\ }\href@noop {} {\bibfield
   {journal} {\bibinfo  {journal} {Physical review letters}\ }\textbf {\bibinfo
  {volume} {113}},\ \bibinfo {pages} {042503} (\bibinfo {year}
  {2014})}\BibitemShut {NoStop}%
\bibitem [{\citenamefont {Bowden}\ \emph {et~al.}(2009)\citenamefont {Bowden},
  \citenamefont {Bernstein}, \citenamefont {Dazeley}, \citenamefont {Svoboda},
  \citenamefont {Misner},\ and\ \citenamefont {Palmer}}]{drift}%
  \BibitemOpen
  \bibfield  {author} {\bibinfo {author} {\bibfnamefont {N.}~\bibnamefont
  {Bowden}}, \bibinfo {author} {\bibfnamefont {A.}~\bibnamefont {Bernstein}},
  \bibinfo {author} {\bibfnamefont {S.}~\bibnamefont {Dazeley}}, \bibinfo
  {author} {\bibfnamefont {R.}~\bibnamefont {Svoboda}}, \bibinfo {author}
  {\bibfnamefont {A.}~\bibnamefont {Misner}}, \ and\ \bibinfo {author}
  {\bibfnamefont {T.}~\bibnamefont {Palmer}},\ }\href@noop {} {\bibfield
  {journal} {\bibinfo  {journal} {Journal of Applied Physics}\ }\textbf
  {\bibinfo {volume} {105}},\ \bibinfo {pages} {064902} (\bibinfo {year}
  {2009})}\BibitemShut {NoStop}%
\end{thebibliography}%
\end{document}